\pdfoutput=1
\documentclass{article}
\usepackage{spconf,amsmath,graphicx}
\usepackage{amssymb}

\usepackage{cleveref}
\usepackage{siunitx}
\usepackage{booktabs}
\usepackage{tikz}
\usetikzlibrary{positioning, fit, calc}
\usetikzlibrary{patterns}
\usepackage{pgfplots}
\usepackage{caption}
\usepackage{csquotes}
\usepackage{balance}
\usepackage{subcaption}
\usepackage{pgfplots}\usepackage[acronym, shortcuts, toc,nonumberlist]{glossaries}
\glsdisablehyper

\usepackage{pifont}

\pgfplotsset{compat=newest}
\usepackage{tabularx, etoolbox, booktabs}
\usepackage{multirow} 

\newcolumntype{H}{>{\setbox0=\hbox\bgroup}c<{\egroup}@{}}  

\newacronym{AAM}{AAM}{additive angular margin}
\newacronym{TAP}{TAP}{Time Average Pooling}
\newacronym{DER}{DER}{Diarization Error Rate}
\newacronym{EER}{EER}{Equal Error Rate}
\newacronym{minDCF}{minDCF}{minimum Detection Cost Function}
\newacronym{EEND}{EEND}{End-to-End Neural Diarization}
\newacronym{VAD}{VAD}{Voice Activity Detection}
\newacronym{PLDA}{PLDA}{Probabilistic Linear Discriminant Analysis}
\newacronym{ASR}{ASR}{Automatic Speech Recognition}
\newacronym{MSE}{MSE}{Mean Squared Error}
\newacronym{TDNN}{TDNN}{Time Delay Neural Network}
\newcommand{\cmark}{\text{\ding{51}}}
\newcommand{\xmark}{\text{\ding{55}}}
\title{Frame-wise and overlap-robust speaker embeddings \\for meeting diarization}
\name{%
\begin{tabular}{@{}c@{}}
Tobias Cord-Landwehr$^{\star}$, %
Christoph Boeddeker$^{\star}$, %
C\u{a}t\u{a}lin Zoril\u{a}$^{\dagger}$, %
Rama Doddipatla$^{\dagger}$,\\%
Reinhold Haeb-Umbach$^{\star}$%
\end{tabular}
}
\address{${}^{\star}$ Paderborn University, Department of Communications Engineering, Paderborn, Germany\\
${}^{\dagger}$ Toshiba Cambridge Research Laboratories, Cambridge, United Kingdom}
\makeatletter
\renewcommand\section{\@startsection {section}{1}{\z@}%
                                   {-2.5ex \@plus -1ex \@minus -.2ex}%
                                   {2.3ex \@plus.2ex}%
                                   {\normalfont\Large\bfseries}}
\renewcommand\subsection{\@startsection{subsection}{2}{\z@}%
                                    {-1.75ex\@plus -1ex \@minus -.2ex}%
                                    {1.5ex \@plus .2ex}%
                                    {\normalfont\large\bfseries}}
\makeatother
\begin{document}
\ninept
\maketitle
\begin{abstract}
Using a Teacher-Student training approach we developed a speaker embedding extraction system that outputs  embeddings at frame rate.
Given this high temporal resolution and the fact that the student produces sensible speaker embeddings even for segments with speech overlap, the frame-wise embeddings serve as an appropriate representation of the input speech signal for an end-to-end neural meeting diarization (EEND) system.
We show in experiments that this representation helps mitigate a well-known problem of EEND systems: when increasing the number of speakers the diarization performance drop is significantly reduced.
We also introduce block-wise processing to be able to diarize arbitrarily long meetings.


\end{abstract}
\begin{keywords}
speaker embeddings, diarization, EEND, d-vectors, teacher-student training, speaker verification
\end{keywords}
\section{Introduction}
\label{sec:intro}
Speaker embedding extraction aims at computing a representation from a speech signal that keeps the information about the speaker identity, while being insensitive to other factors, such as the content.
Next to its obvious application in speaker recognition or verification tasks, there are 
other applications that benefit from such a voice fingerprint, such as \gls{ASR} \cite{Saon13ivectorASR}, source extraction from a speech mixture \cite{Zmolikova19SpeakerBeam} and speaker diarization \cite{ryant21dihard3, watanabe20b_chime}.

Early approaches to speaker embedding computation (e.g., the i-vectors \cite{11_dehak_ivectors}) employed statistical methods, however, more recently, the neural network-based techniques are prevalent. 
In the latter case, the speaker embedding is taken as the representation at a bottleneck layer 
right before the last, i.e., the classification layer of the network. 
X-vectors \cite{Snyder18xvectors} are a popular neural network based representation that employs a \gls{TDNN} with statistics pooling for computation. Another widely used speaker embeddings are the d-vectors that use ResNet-based convolutions for computation \cite{DBLP:journals/corr/LiMJLZLCKZ17}, achieving an impressive speaker verification performance on  VoxCeleb \cite{Nagrani19VoxCeleb}.


Speaker embeddings have also been used for speaker diarization. For example, the baseline system of the third DiHARD challenge \cite{ryant21dihard3} employed an x-vector extractor, followed by similarity scoring and clustering. 
However, there are at least two issues when used in speaker diarization tasks. Firstly, the embeddings are computed on a relatively large signal window  of typically \SIrange{2}{4}{\second}. This contradicts the benefit to have a high temporal resolution of the diarization information of a few frames. Secondly, embeddings computed from regions of overlapped speech, i.e., regions where more than one speaker is active, are typically uninformative. Ideally,
the computed embedding vector for a mixture of two speakers should  also be a superposition of the embeddings of the participating speakers. However, the embedding extraction is far from being linear, and the participating speakers cannot easily be inferred from 
the speech mixture.
This is a well-known issue that hurts clustering-based approaches to speaker diarization.

In contrast, the \gls{EEND} \cite{fujita19_interspeech} method can handle speech overlap regions well by formulating diarization as a multi-label classification problem. However, \gls{EEND} systems suffer from the fact that 
they solve this problem rather \enquote{locally},
meaning that they can
discern only a small number of speakers
and 
they struggle to re-identify the speakers at a later point in time.
To alleviate these issues, methods like EEND-EDA compute speaker-wise attractors
to maintain a speaker state \cite{Horiguchi22EDAEEND, Horiguchi21localglobalEDAEEND}, 
or employ \gls{ASR}-generated features to achieve a similar goal \cite{Khare22ASREEND}.

Rather than fixing the problem posthoc by modifying the architecture of neural diarization, in this contribution we directly address the above two deficiencies of state-of-the-art speaker embedding extractors. This will in turn improve the speaker re-identification when using the embeddings as input to a neural diarization system.

Our proposal is to decouple the learning of the speaker embedding space from the projection of a speech signal into that space, using a Teacher-Student strategy. The teacher is a standard ResNet d-vector system, and the student is trained to output high temporal resolution (i.e. frame-level) vectors that replicate the teacher's embeddings. 
Therefore, the student projects a segment of input speech to the embedding vector representing the active speaker. 
While a pooling over several seconds is crucial for the teacher to learn a robust speaker embedding space, the student can work at a much higher temporal resolution.
This leads to some performance drop in a large scale speaker verification task, but works sufficiently well for small numbers of speakers, as is typical for meeting diarization scenarios.
Additionally, the frame-level resolution has the benefit that simultaneously active speakers are better recognizable from the embeddings computed from a mixture, because many frames are dominated by either of the two speakers. The pooling of the teacher, on the other hand, would average out the speaker-specific information in the mixture.
Using the student embedding extractor as a front-end to a \gls{EEND} system, we show that diarization performance is significantly improved on meeting data compared to the baseline \gls{EEND} system.

In the next section we describe the architectures of the teacher and the student network as well as the Teacher-Student training. In \Cref{sec:student_eend} we motivate the usage of the student-embeddings as input features for neural speaker diarization. Then, the quality of the frame-wise student embeddings, both for single-speaker data and two-speaker mixtures, as well as their applicability for diarization, is evaluated in \Cref{sec:evaluation}.

\section{Teacher-Student Embeddings}
\label{sec:ts_embeddings}
The extraction of the speaker characteristics from an input speech signal using a neural network is typically achieved by introducing an information bottleneck right before the classification layer of the network. The network is trained to minimize a classification loss between the estimated and the true speaker label.
Since the number of speakers seen during training is significantly larger than the dimensionality of the latent embedding space, the representation at the bottleneck layer is forced to represent the speaker characteristics and  not only the speaker labels. In this way, speaker embedding extractors  generalize well to unseen speakers \cite{Snyder18xvectors}.

Typically, a single embedding is computed from an utterance or from a few seconds of speech. This is done by \gls{TAP} of the activations of the bottleneck layer. In fact, it has been shown that this aggregation over several seconds is necessary to achieve a good classification performance \cite{gusev20_odyssey}.

When employing these systems for diarization, the utterance is split into  overlapping windows of 2 to 4 seconds
length, from which one embedding vector per window is computed.
However, the \gls{TAP} renders the speaker embeddings unreliable if the environment changes over the duration of a window \cite{22LandiniVBx}.
In this case, the aggregation of the activations does not deliver a meaningful speaker representation, as their properties change midway.
This is a well-known issue of classical clustering-based diarization approaches \cite{22LandiniVBx, Doverlap}.

In order to obtain a speaker embedding extractor that does not require \gls{TAP} and still computes reliable embeddings, we employ a Teacher-Student approach. The teacher is an ordinary speaker embedding extractor, and the student is the desired extractor that can produce reliable frame-wise embeddings, as is described next.

\subsection{Teacher speaker embeddings}
\label{sec:teacher}
In this work, the teacher network is a typical ResNet-based d-vector extractor \cite{Zhou21ResNet}, that is widely used for speaker verification \cite{Qin22VoxSRC22DKU, Suh22Voxsrc22}.
Here, first logarithmic mel filterbank features $x(t,f)$ are extracted from an observation $x(\ell)$. These features are then encoded into frame-wise latent embeddings, which are then aggregated with a \gls{TAP} layer to obtain a single, $E$-dimensional d-vector $d(e)$ representing the speaker characteristics of the input speech.
During training, an additional fully connected classification layer is used to predict the speaker label.
By using a softmax-based classification loss that normalizes the embeddings to unit length, the length-normalized d-vectors lie on a $E$-dimensional hypersphere 
once the model is fully converged.
However, this property only holds true for the d-vectors $d(e)$, not the frame-wise embeddings before the \gls{TAP}.
Other works \cite{Kwon21NotTalking}  showed that these frame-wise embeddings heavily fluctuate in terms of power and can even be used to infer a \gls{VAD} from, so that no further assumptions about the structure of this data can be made.

\subsection{Teacher-Student training}
\label{sec:student}
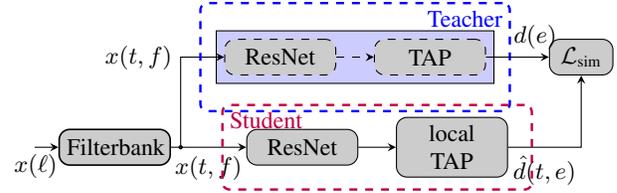
\begin{figure}
    \centering

\begin{tikzpicture}[semithick,auto,
block_high/.style={
		rectangle,
		draw,
		fill=black!20,
		text centered,
		text width=4em,
		rounded corners,
 		minimum height=1.5em,
		minimum width=2em},
block/.style={
	rectangle,
	draw,
	fill=black!20,
	text centered,
	text width=4em,
	rounded corners,
	minimum height=1.5em,
	minimum width=2em},
block_loss/.style={
	rectangle,
	draw,
	fill=black!20,
	text centered,
	text width=2em,
	rounded corners,
	minimum height=1.5em,
	minimum width=2em},
block_teacher/.style={
		rectangle,
		draw,
		dashed,
		fill=black!20,
		text centered,
		text width=4em,
		rounded corners,
 		minimum height=1.5em,
		minimum width=3em},
mul/.style={
        circle,
        draw,
    },		
	]
\tikzset{>=stealth}
\tikzstyle{branch}=[{circle,inner sep=0pt,minimum size=0.3em,fill=black}]

\tikzset{pics/.cd,
	pic switch closer/.style args={#1 times #2}{code={
		\tikzset{x=#1/2,y=#2/2}
		\coordinate (-in) at (1,0);
		\coordinate (-out) at (-1,0);
		
		\draw [line cap=rect] (-1, 0) -- ++(0.1,0) -- ++(20:1.9);
		
	}}
}

    \pgfdeclarelayer{background1}
    \pgfdeclarelayer{background2}
    \pgfdeclarelayer{mid1}
    \pgfdeclarelayer{mid2}
    \pgfdeclarelayer{foreground}
    \pgfsetlayers{background1,background2,mid1,mid2,main,foreground}

    \node[block_high] (encoder) {Filterbank};
    \node[circle, draw=black, inner sep=0cm, right=0.1cm of encoder] (split_enc) {};
	\begin{pgfonlayer}{background2}
		\node [block_teacher, above right=1cm of encoder] (teacherbg) {ResNet};
		\node [block_teacher, right=0.5cm of teacherbg] (poolingbg) {TAP};
	\end{pgfonlayer}
	\begin{pgfonlayer}{background1}
		\node [draw=black, fill=blue!20, fit={(teacherbg) (poolingbg)}] {};
	\end{pgfonlayer}

	    \tikzstyle{box} = [draw, dashed, inner xsep=1em, inner ysep=1.5em, line width=0.1em, rounded corners=0.3em]
		\node [box, draw=blue,fit={(teacherbg) (poolingbg) }, label={[anchor=north east, align=left]north east:\color{blue}Teacher}] {};
    \node [block, right=1cm of encoder] (student) {ResNet};
    \node [coordinate, right=3.5cm of student] (splitframe) {};

    \node [block_loss] (loss) at ($(poolingbg.east) + (4em,0)$) {$\mathcal{L_{\text{sim}}}$};
    \node[coordinate, right=0.5cm of encoder] (splitteacher) {};
    \node[block, right=0.5cm of student] (shortpooling) {local TAP};
   \draw[<-] ($(encoder.west)$) node[xshift=-0.3cm, yshift=-0.25cm] {$x(\ell)$} -- +(-1em, 0em);

   \begin{pgfonlayer}{background1}
   	\draw[->] (split_enc) |- node[midway,left] {$x(t,f)$}  ($(teacherbg.west) + (-0em,0)$);
   \end{pgfonlayer}

   \draw[->] ($(encoder.east)$) --node[below] {$x(t,f)$} (student);
   \draw[->] ($(poolingbg.east) +(0.35em,0)$) -- node[near end,above] {$d(e)$}($(loss.west) +(0,0)$);

  \draw[->, dashed] (teacherbg) -- (poolingbg);

 \draw[-, dashed] (poolingbg) -- ($(poolingbg.east) +(0.35em,0)$);

  \tikzstyle{box} = [draw, dashed, inner xsep=1em, inner ysep=0.5em, line width=0.1em, rounded corners=0.3em]
  \node [box, draw=purple,fit={(student) (shortpooling) }, label={[anchor=north west, align=right]north west:\color{purple}Student}] {};
   \draw[->] ($(student.east)$) |-  node[near end, below] {} ($(shortpooling.west)$);
   %

   %
   \draw[->] ($(shortpooling.east)$) -|  node[near start, below] {$ \hat d(t,e)$}($(loss.south) +(0em,0)$);

\end{tikzpicture}
    \caption{Block diagram of the student training. The teacher d-vectors serve as targets for the frame-wise embeddings of the student.}
    \label{fig:teacher_student}
\end{figure}

For the Teacher-Student training, a second network, the student, is trained while using the d-vectors $d(e)$ of the teacher as training targets as depicted in \Cref{fig:teacher_student}. The weights of the teacher itself are kept fixed during this step.
The architecture of the student network is largely identical to the teacher.
Here, also a ResNet34 extracts frame-wise speaker embeddings from the logarithmic mel  filterbank features $x(t,f)$ of the observation.

However, the student exhibits two key differences to the teacher.
First, instead of performing a global \gls{TAP}, a local \gls{TAP} over \num{11} frames with a frame advance of single frame to smooth the output embeddings is applied. In this way, the student embeddings $\hat{d}(t,e)$  maintain a frame-wise resolution, while they still include the contextual frames during embedding extraction.
Second, the student is not trained with a classification loss, but with a \gls{MSE}-based similarity loss
\begin{align}
    \mathcal{L}_{\mathrm{sim}} = \frac{1}{T E} \sum_{t=1}^T \sum_{e=1}^E \lVert d(e) - \hat{d}(t,e) \rVert^2
\end{align}
between each frame-wise student embedding $\hat{d}(t,e)$ and the teacher d-vector $d(e)$.
Through this loss, the student is not only encouraged to reproduce the d-vectors of the teacher, but to do so on a frame level.
Therefore, every frame-wise student embedding ideally is projected onto the $E$-dimensional hypersphere that depicts the latent speaker space of the teacher. 
Those  are easier to interpret, in particular in case of a speaker change or speech overlap, since they depict a higher time resolution.

\section{Speaker embedding-based EEND}
\label{sec:student_eend}
To evaluate the previous hypothesis, we adapt the \gls{EEND} approach \cite{fujita19_interspeech}.
In  \gls{EEND} systems, a neural network is trained to directly predict the activity of each speaker in a meeting recording from the observation.
The \gls{EEND} model consists of a feature extraction followed by self-attention layers and a classification layer to predict the frame-wise activity of each speaker.
During inference, an additional min pooling (erosion) followed by a max pooling (dilation) is applied before thresholding the model output to obtain the estimated activities. 

\begin{figure}[bt]
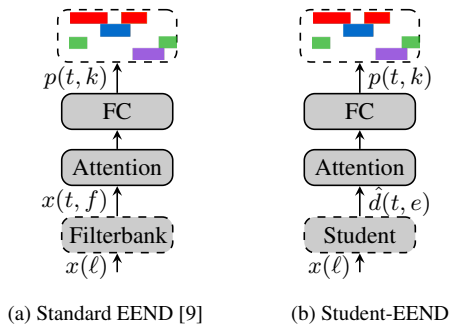

     \centering
     \subcaptionbox{Standard EEND \cite{fujita19_interspeech}}[.4\columnwidth]{\input{tikz/EEND}}
     \subcaptionbox{Student-EEND}[.4\columnwidth]{\input{tikz/EEND_student}}
     \caption{Comparison of the standard EEND \cite{fujita19_interspeech} and Student-EEND}
     \label{fig:student_eend}
 \end{figure}

Here, we replace the feature extraction with the student embedding extractor as depicted in \Cref{fig:student_eend} and train the \gls{EEND} components while keeping the model weights of the student frozen. The resulting system is denoted \enquote{Student-EEND}.
Since the speaker information is already encoded in the input features to the \gls{EEND} model, the Student-EEND only needs to learn a multi label assignment of these input features to provide a good diarization, and the speaker information necessary to distinguish between speakers is directly available at the input.

\subsection{Block-wise Student-EEND}
For normal \gls{EEND} models, the full length audio of the recording is required at once.
In order to obtain a diarization system that can operate on arbitrarily long meetings, a block-wise processing is proposed in the following. 
Instead of processing the whole meeting at once, during inference the meeting is processed in overlapping blocks.
Note that block processing results in a block permutation problem, i.e.,
the estimated activity $\bar{p}_b(t,k)$ after thresholding in each block $b$ can be arbitrarily permuted.
Block-wise processing has also been proposed in  \cite{Kinoshita21Intregration, xue21d_interspeech}. In these works, the models are trained to output an additional state that can be used to solve the block permutation problem.
Here, instead of inferring a speaker state from the \gls{EEND} output, we directly use the student speaker embeddings to reorder the activity estimates and assign them to the correct speaker.
Similarly to \cite{Kinoshita21Intregration}, this is done by first obtaining a block-wise d-vector by computing a weighted mean
\begin{align}
d_b(k,e) = \frac{1}{T} \sum_{t=1}^T \bar{p}_b(t,k)\hat{d}_b(t,e)
\end{align}
for each active speaker $k$ in the processed block $b$ from the student embeddings $\hat{d}_b(t, e)$  with the estimated activities $\bar{p}_b(k,t)$. 
Then, inactive speakers are removed and the d-vectors of all blocks are clustered with an overlap-aware k-means clustering to find the block permutation.
Next,  
the activities 
in the center parts 
of successive blocks 
are concatenated as in \cite{20_Chen_LibriCSS} to obtain the diarization estimate.
The number of speakers active in a block 
tends to be smaller than the total number of speakers participating in a meeting. Therefore, the Student-EEND system can be trained for an expected maximal number of speakers in a block (in this work, maximum 4 speakers) and then applied to a meeting with arbitrarily many speakers.
In this way, the Student-EEND can be used for a block-wise processing without any further modifications to the 
\gls{EEND} model.

\section{Evaluation}
\label{sec:evaluation}
\subsection{Databases and training setup}
The teacher and student models are trained using VoxCeleb data \cite{Nagrani19VoxCeleb} augmented with noise from the MUSAN dataset \cite{snyder2015musan} and RIRs from \cite{Drude_19_smswsj}.
The teacher is trained with the \gls{AAM} Softmax \cite{liu19f_interspeech} loss and the student is trained with a frame-wise \gls{MSE} loss to match the teacher's d-vectors.
%
Reverberant LibriSpeech meetings are
simulated with the MMS-MSG software \cite{cordlandwehr2022mms_msg} which allows for the simulation of meetings with an arbitrary number of speakers and duration. These meetings contain \SIrange{15}{20}{\percent} overlap and are used to evaluate the speaker embeddings under overlapping speech.
While the teacher and student systems are only evaluated on the test meetings, the \gls{EEND} systems also use matched data for training.
\SI{2}{\minute} meetings are simulated and all the diarization models use \SI{60}{\second} chunks for training.
During evaluation, the \gls{EER} and \gls{minDCF} are calculated or speaker verification  and the \gls{DER} for diarization performance according to \cite{nist}.


\subsection{Student embeddings for speaker verification}
For the speaker verification performance, both the student and the teacher d-vector models are evaluated on the VoxCeleb speaker verification test sets.
In this context, for better comparison with the teacher, the frame-wise embeddings $\hat d(t,e)$ of the student also are averaged over time to obtain a single student d-vector $\bar{d}(e)$.  
As can be seen in \Cref{tab:eer_vc} ,
the student d-vectors consistently perform worse than the teacher d-vectors for this task.
Finetuning the student model training by replacing the target d-vector with a different d-vector of the same speaker reduces the performance gap to some degree.
\begin{table}[bt]
\setlength{\tabcolsep}{4pt}
    \centering
    \renewcommand{\arraystretch}{0.95}
    \caption{Speaker verification performance of the teacher and student models on the normal (O) and hard (H) VoxCeleb trial sets.}
        \sisetup{detect-weight}
	\robustify\bfseries  
    \sisetup{round-precision=2,round-mode=places, table-format = 2.1}
    \centering
    \begin{tabular}{l S S  S S}
    \toprule
        Model  & \multicolumn{2}{c}{{VoxCeleb1-O}} & \multicolumn{2}{c}{{VoxCeleb1-H}} \\
        \cmidrule(lr){2-3}\cmidrule(lr){4-5}
        & {EER[\si{\percent}]} & {minDCF} & {EER[\si{\percent}]} & {minDCF} \\
        \midrule
        ECAPA-TDNN \cite{Brecht20EscapaDNN} & 0.87 & 0.07 & 2.12 & 0.13\\
        \midrule
        Teacher &  1.06 &  0.08 &  2.71 &  0.15\\
        Student & 2.34 & 0.19 & 5.33 & 0.29 \\
        + finetuning   & 2.02 & 0.14 & 4.16 & 0.21 \\
         \bottomrule
    \end{tabular}

    \label{tab:eer_vc}
\end{table}
The worse performance of the student is not surprising, as only few student models are able to outperform their teachers on their original tasks \cite{hinton2015distilling}.
The student is, however, expected to perform reasonably well with a small number of speakers, as is the case for meetings.



\subsection{Embeddings on overlapped speech}
\label{sec:consistency}
For the purpose of conversational speech diarization, an important question is whether the embeddings are robust under overlapping speech. For this purpose, the teacher and student speaker embedding extractors are evaluated for 8-speaker LibriSpeech meetings.
Both for the teacher and the student, short-time d-vectors with a duration of \SI{2}{\second} and a hop size of \SI{0.5}{\second} are extracted from these data.
The lengths correspond to sizes typically chosen for speaker embedding based diarization systems. 
Then, prototype d-vectors for each speaker in the meeting are obtained from clean  recordings, and they serve as a reference for the  above segment d-vectors to identify the active speakers both in the single-speaker and overlapping regions.
This is done  by computing the cosine similarity between all prototypes and the segment d-vector. For overlap, the d-vectors are assigned to the two most similar speakers.
\Cref{tab:ov_assignment}  shows that student and teacher achieve the same accuracy for the identification of single speaker regions.
As expected, in the overlap regions, however, the student's identification accuracy outperforms the teacher, indicating a higher robustness for overlapping speech. 
\begin{table}
    \renewcommand{\arraystretch}{0.95}
    \centering
    \caption{Speaker identification accuracy of teacher and student d-vectors for single speaker and overlap regions for LibriSpeech meetings}
    \sisetup{detect-weight}
	\robustify\bfseries  
    \sisetup{round-precision=2,round-mode=places, table-format = 2.1}
    \begin{tabular}{l S S}
    \toprule
    Model    & {Single speaker}&  {Overlap} \\
    \midrule
    Teacher & 0.9774754816037197 & 0.35198948708312376 
 \\
    Student & 0.9777148764521461 & \bfseries 0.42 \\
    \bottomrule
    \end{tabular}
    \label{tab:ov_assignment}
\end{table}

\begin{figure}[bt]
    \centering
    \input{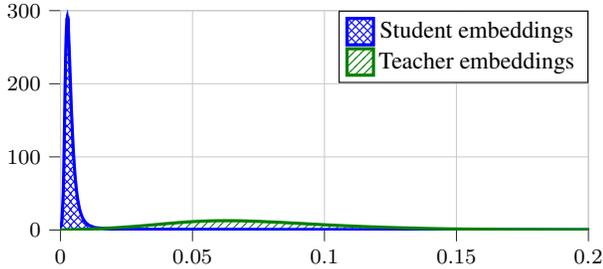}
    \caption{Probability density of local cosine distance between consecutive frame-wise embeddings for reverberant 8-speaker meetings}
    \label{fig:local_dist}
\end{figure}

\Cref{fig:local_dist} explains why the student embeddings are able to better represent the active speakers for overlapping speech over the teacher embeddings.
The figure displays the accumulated local cosine distance between successive frame-wise speaker embeddings.
As shown, the distances for the student are much smaller than for the teacher. 
This indicates that the student's frame-wise embeddings smoothly change from one speaker to the next even during the overlap.
We further observed that higher local distances coincide with speaker changes over the course of a meeting for the student.

\begin{figure}[bt]
\hspace{1em}
    \subcaptionbox{Teacher embeddings}{\input{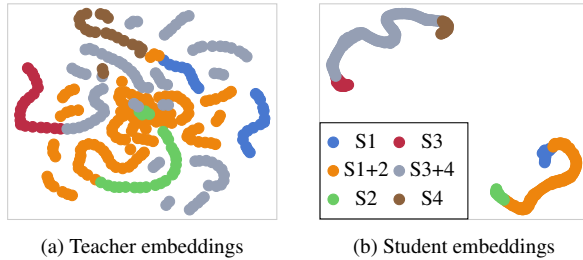}}
    \hspace{1em}
    \subcaptionbox{Student embeddings}{

\begin{tikzpicture}

\definecolor{color0}{rgb}{0.282352941176471,0.470588235294118,0.815686274509804}
\definecolor{color1}{rgb}{0.933333333333333,0.52156862745098,0.050196078431373}
\definecolor{color2}{rgb}{0.415686274509804,0.8,0.392156862745098}
\definecolor{color3}{rgb}{0.73921568627451,0.172549019607843,0.272549019607843}
\definecolor{color4}{rgb}{0.584313725490196,0.623529411764706,0.705882352941177}
\definecolor{color5}{rgb}{0.549019607843137,0.380392156862745,0.235294117647059}

\begin{axis}[
axis line style={white!80!black},
xmajorticks=false,
xmin=-23.9594908714294, xmax=23.0571986198425,
ymajorticks=false,
ymin=-23.6421216011047, ymax=22.1087563514709,
legend style={at={(0,0)}, anchor=south west, font=\footnotesize, legend columns=2},
width=0.6\columnwidth
]
\addplot [draw=color0, fill=color0, mark=*, only marks]
table{%
x  y
15.3658638000488 -11.5483703613281
15.4049320220947 -11.2871627807617
15.4426183700562 -10.9645214080811
15.4724683761597 -10.7112550735474
15.5395088195801 -10.2615184783936
15.391077041626 -9.89814281463623
14.9901161193848 -9.60240840911865
15.0692148208618 -9.56655597686768
15.4294013977051 -9.47980213165283
15.2788324356079 -9.14371299743652
15.398232460022 -8.95605373382568
15.9794445037842 -9.18571186065674
16.2325134277344 -9.00273609161377
16.3015670776367 -8.5773754119873
16.6143074035645 -8.52310562133789
};
\addlegendentry{S1}
\addplot [draw=color3, fill=color3, mark=*, only marks]
table{%
x  y
-18.9324703216553 4.89883422851562
-19.0980281829834 4.90731859207153
-19.277156829834 4.89574956893921
-19.5017719268799 4.63173627853394
-19.6126365661621 5.04458475112915
-19.950309753418 4.6418194770813
-20.1844749450684 4.67360210418701
-20.4745426177979 4.76249122619629
-20.2069339752197 5.2900767326355
-20.681713104248 5.09020042419434
-20.3384170532227 5.6369194984436
-20.9556674957275 5.44230270385742
-20.9272422790527 5.75561332702637
-20.7367134094238 6.16883611679077
-21.0636901855469 6.3763484954834
};
\addlegendentry{S3}

\addplot [draw=color1, fill=color1, mark=*, only marks]
table{%
x  y
16.8544750213623 -8.21671867370605
17.003023147583 -7.78724718093872
17.1551647186279 -7.56600189208984
17.4443817138672 -7.62302780151367
17.7630729675293 -7.7151050567627
18.0118083953857 -7.43248081207275
18.2644176483154 -7.46477842330933
18.4752902984619 -7.83221197128296
18.8899097442627 -7.64142751693726
19.0545253753662 -7.77410554885864
19.1424083709717 -8.09177398681641
19.2404899597168 -8.42091178894043
19.6800785064697 -8.50151062011719
19.7356472015381 -8.76883125305176
19.922212600708 -9.11298561096191
20.1258850097656 -9.46561622619629
20.3998756408691 -9.66277599334717
20.4339122772217 -9.97336387634277
20.5154685974121 -10.2889270782471
20.7243976593018 -10.6341037750244
20.8461132049561 -10.8826513290405
20.8197078704834 -11.2111148834229
20.9200763702393 -11.5707607269287
20.8994274139404 -11.9698858261108
20.8616256713867 -12.2757949829102
20.8014240264893 -12.7152643203735
20.6559219360352 -13.1428213119507
20.5261764526367 -13.4851217269897
20.3758850097656 -13.7661581039429
20.2565021514893 -14.0776233673096
19.9027061462402 -14.2686710357666
19.9194374084473 -14.6754169464111
19.4880466461182 -14.5556058883667
19.4125366210938 -15.0033779144287
18.97145652771 -14.8484001159668
18.941614151001 -15.2614917755127
18.6182765960693 -15.4581489562988
18.261209487915 -15.2657670974731
18.0342025756836 -15.2636156082153
17.8531074523926 -15.7088422775269
17.4186458587646 -15.4602661132812
17.2936534881592 -15.8265180587769
16.8096809387207 -15.6490411758423
16.7712116241455 -16.050760269165
16.5383529663086 -16.1806182861328
16.1768550872803 -16.1128368377686
15.8662662506104 -16.3913726806641
15.5876150131226 -16.624361038208
15.3037328720093 -16.8455562591553
14.9972867965698 -17.0549964904785
14.6593523025513 -17.3849430084229
14.3080806732178 -17.6929035186768
14.0942525863647 -17.9619903564453
13.8220911026001 -18.0506725311279
13.8432884216309 -18.3918647766113
13.4157180786133 -18.386531829834
13.4805850982666 -18.7459564208984
13.171236038208 -18.9148559570312
13.1879787445068 -19.3736476898193
12.8065948486328 -19.4116477966309
12.9885034561157 -19.8215999603271
12.6397705078125 -19.9908180236816
12.6133146286011 -20.3690090179443
12.6143827438354 -20.7535858154297
12.2179384231567 -20.7523899078369
12.2816781997681 -21.2058544158936
12.0931873321533 -21.4257583618164
11.7731895446777 -21.3517265319824
11.5269842147827 -21.562536239624
11.2058172225952 -21.1772975921631
10.9364891052246 -21.4435901641846
10.6555938720703 -21.1743068695068
10.3605403900146 -21.0446662902832
10.1271352767944 -20.7261028289795
9.76595687866211 -20.7082347869873
9.54086875915527 -20.4555130004883
9.21974849700928 -20.2511882781982
9.0474214553833 -19.9980392456055
8.72768592834473 -19.8864994049072
8.55919551849365 -19.5699119567871
};
\addlegendentry{S1+2}
\addplot [draw=color4, fill=color4, mark=*, only marks]
table{%
x  y
-21.1244277954102 6.7946982383728
-21.2552757263184 7.17888164520264
-21.3755531311035 7.58651781082153
-21.4901428222656 7.99546194076538
-21.5465698242188 8.27472114562988
-21.7078456878662 8.56864833831787
-21.6242046356201 8.8516149520874
-21.8223686218262 9.15707397460938
-21.519193649292 9.38778018951416
-21.6828422546387 9.69625186920166
-21.4933471679688 9.96658229827881
-21.3657741546631 10.2514095306396
-21.262903213501 10.564359664917
-21.147762298584 10.8965835571289
-21.0029621124268 11.1988134384155
-20.8805484771729 11.5050506591797
-20.7010459899902 11.8679914474487
-20.5420989990234 12.1706418991089
-20.3280410766602 12.6054515838623
-20.1376399993896 13.0181446075439
-19.800952911377 13.2126893997192
-19.8840503692627 13.644907951355
-19.7937641143799 13.9396018981934
-19.263090133667 13.8305606842041
-19.5726299285889 14.3991212844849
-19.1341514587402 14.2158985137939
-18.8068237304688 14.2880020141602
-19.0529003143311 14.8473510742188
-18.8468742370605 14.9526109695435
-18.3631534576416 14.7967290878296
-18.094913482666 14.8800086975098
-18.100492477417 15.4008226394653
-17.7622127532959 15.533447265625
-17.3866367340088 15.3601913452148
-17.286600112915 15.8885536193848
-16.922643661499 15.6684465408325
-16.7467765808105 15.9622707366943
-16.410285949707 15.9226684570312
-16.1487731933594 15.761589050293
-15.9368190765381 15.6027336120605
-15.6810874938965 15.3569078445435
-15.3658800125122 15.1378288269043
-14.9756517410278 14.8934183120728
-14.6236333847046 14.6753206253052
-14.2894315719604 14.4735012054443
-13.9839191436768 14.2974214553833
-13.6408376693726 14.1946144104004
-13.2114019393921 14.0956230163574
-12.7492179870605 13.9479856491089
-12.5337409973145 13.8747043609619
-12.2488317489624 14.0160846710205
-11.9234895706177 13.9018783569336
-11.5795412063599 13.7751693725586
-11.4144878387451 13.96875
-11.4092378616333 14.3487339019775
-10.9576263427734 14.2986688613892
-10.8206491470337 14.5012245178223
-10.956127166748 14.892370223999
-10.7327260971069 15.1824989318848
-10.6782178878784 15.5439653396606
-10.7376537322998 15.7650270462036
-10.8340120315552 16.0421447753906
-11.0087699890137 16.2626438140869
-11.0817604064941 16.5738925933838
-11.3048477172852 16.8066844940186
-11.4765882492065 17.0234985351562
-11.6974773406982 17.3853435516357
-11.9243583679199 17.7641906738281
-12.1295766830444 18.1878185272217
-12.251522064209 18.5079345703125
-12.2490072250366 18.8044548034668
-12.2025985717773 19.13450050354
-12.0219202041626 19.3562793731689
-11.8115377426147 19.5176258087158
-11.3927659988403 19.7127590179443
-10.9498977661133 19.8745765686035
-10.5457048416138 19.9867935180664
-10.0999526977539 20.0068168640137
-9.53031253814697 20.0252799987793
-8.94010448455811 20.0291709899902
-8.57175254821777 20.0003318786621
-8.23879241943359 19.9456577301025
-7.8828182220459 19.886381149292
-7.6329984664917 19.8571891784668
-7.44427394866943 19.7684745788574
-7.07076930999756 19.5938186645508
-6.72000312805176 19.4220199584961
-6.37212610244751 19.2639122009277
-6.1805419921875 19.1191234588623
-5.88852643966675 18.9649391174316
-5.54468727111816 18.9088535308838
-5.39451551437378 18.6573810577393
-5.14072608947754 18.7264957427979
-4.87758302688599 18.6445980072021
-4.51936435699463 18.6810245513916
-4.19702625274658 18.6261405944824
-3.87658166885376 18.5430355072021
-3.47306823730469 18.5205497741699
-3.14488816261292 18.4264602661133
-2.85156893730164 18.3970832824707
};
\addlegendentry{S3+4}
\addplot [draw=color2, fill=color2, mark=*, only marks]
table{%
x  y
8.25748062133789 -19.4878311157227
8.1503267288208 -19.1897716522217
7.79021120071411 -19.0952281951904
7.66203546524048 -18.7916812896729
7.33258628845215 -18.6395301818848
7.37865018844604 -18.2248344421387
6.96605253219604 -18.176118850708
6.83265733718872 -17.96506690979
6.87048625946045 -17.7065963745117
6.91174364089966 -17.4983921051025
6.96554565429688 -17.3574256896973
6.93831205368042 -17.1912574768066
};
\addlegendentry{S2}

\addplot [draw=color5, fill=color5, mark=*, only marks]
table{%
x  y
-2.45098042488098 18.4276084899902
-2.3405704498291 18.1588230133057
-1.97410774230957 18.2002792358398
-1.91812098026276 17.8813095092773
-1.61100733280182 17.7392864227295
-1.4949666261673 17.4314918518066
-1.79798340797424 17.1485042572021
-1.5291713476181 16.8856525421143
-1.62106609344482 16.5772571563721
-2.10568404197693 16.552267074585
-1.95570421218872 16.2161388397217
-2.19427609443665 16.0033683776855
-2.39158391952515 15.9779510498047
-2.58793234825134 15.8692235946655
-2.7235369682312 15.7319316864014
};
\addlegendentry{S4}
\end{axis}

\end{tikzpicture}}
    \caption{t-SNE plot for the frame-wise embeddings of two segments consisting of partially overlapping speech.}
    \label{fig:tsne}
\end{figure}

This smooth transition can also be seen when visualizing the frame-wise embeddings with t-SNE. \Cref{fig:tsne} shows that, for the student, the embeddings computed from overlapped speech segments form trajectories connecting the embeddings computed from single-speaker regions of the  speakers present in the mixture, whereas no discernible geometry can be seen for the teacher embeddings. 
Enforcing the student vectors to lie in the latent space defined by the teacher seems to help regularize the frame-wise speaker embeddings for overlapping speech, even if the student was not trained on speech mixtures.
Note, that it cannot be concluded that these
embeddings are linear combinations of the participating speaker embeddings, because t-SNE is a highly nonlinear mapping from a high-dimensional space to two dimensions.

\subsection{Embeddings for diarization}
\label{sec:eend}

The previous evaluations show that the frame-wise student embeddings are meaningful even under speech overlap.
While simply assigning regions of overlapping speakers to the most similar speaker prototypes is still too unreliable, it should be possible to find a nonlinear mapping of the frame-wise student embeddings which leads to an accurate diarization.
Therefore, we use the Student-EEND model to this end and compare it to the standard \gls{EEND}.
The values for the activity threshold, erosion and dilation are finetuned separately for each system configuration.
The training is performed using a permutation invariant binary cross entropy loss.

\Cref{tab:EEND_vs_stud_EEND} shows that the student embeddings indeed can be used for diarization.
For meetings with two speakers, both the standard \gls{EEND} and Student-EEND perform similarly.
However, when switching to meetings with four active speakers, the standard \gls{EEND} steeply degrades, while the student embedding-based version still is able to provide a decent diarization result and clearly outperforms the \gls{EEND} model. 
To verify that it is really the Teacher-Student training that is responsible for this effect, we also train an \gls{EEND} model directly on the activations of the teacher d-vector system, i.e., on the embeddings before \gls{TAP}, which we denote \enquote{Teacher-EEND}. As shown, this configuration is unable to provide a meaningful diarization even for the two-speaker case.
This again demonstrates that the student embeddings are easier to interpret on a frame-level and that they form a useful front-end for a neural diarization system.

\subsection{Block-wise Student-EEND}
Next, we investigate the performance of the Student-EEND for a block-wise processing. 
\Cref{tab:blockwise_EEND} compares the block-wise approach with processing the full 10-mins meetings at once.
The block-wise system is trained on 4-speaker meetings and evaluated on \SI{30}{\second} long blocks with a block advance of \SI{10}{\second}.
The results for processing the full meeting have been obtained with systems that have been trained specifically for the given number of speakers. 
It can be seen that the block processing leads to an increase in the \gls{DER}, which is mainly due to the fact that the clustering could not resolve all block permutation errors. Furthermore, it can be observed that the least relative increase in DER is observed if the number of speakers in training matches those in test (i.e., 4). While the generalization to a smaller number of speakers is satisfactory, the deterioration for 8-speaker system to \SI{31}{\percent} is more pronounced. This observed performance loss for more speakers is comparable to the degradation reported in  \cite{kinoshita2022tight}.
Notably, training a 8-speaker \gls{EEND} model was not possible due to the complexity associated with solving the permutation problem.



\begin{table}[bt]
    \renewcommand{\arraystretch}{0.95}
    \centering
    \caption{DER (Missed Hit/ False Alarm/ Confusion) on simulated LibriSpeech meetings with a duration of \SI{2}{\min}}
    \label{tab:EEND_vs_stud_EEND}
    \begin{tabular}{l c c}
    \toprule
        Model &  {2 speakers} & {4 speakers}\\
        \midrule
        EEND &  {\bf 6.1} (2.1 / 2.6 / 1.4) & 33.2 (21.4 / 3.3 / 8.5)\\
        Student-EEND & 6.6 (2.4 / 1.9 / 2.3) & {\bf 13.4} (6.9/ 3.9 / 2.6)\\
        Teacher-EEND &17.9 (6.0 / 8.5/ 3.4) & 49.6 (0.4 / 32.9 / 16.3) \\
        \bottomrule
    \end{tabular}
\end{table}

\begin{table}[bt]
    \renewcommand{\arraystretch}{0.95}
    \sisetup{detect-weight}
	\robustify\bfseries  
    \centering
    \caption{DER for \SI{10}{\minute} LibriSpeech meetings with and without block-wise processing}
    \label{tab:blockwise_EEND}
    \begin{tabular}{l c c c c}
    \toprule
        Model & block-wise &\multicolumn{3}{c}{\#Speakers per Meeting}\\
        &  &{2} & {4} & {8}\\
        \midrule
        EEND & \xmark &\bfseries 6.5& 38.8& -\\
        Student-EEND  & \xmark&   7.6  &  \bfseries 14.4 & -\\
        Student-EEND (4spk) & \cmark & 9.6 &16.8 & \bfseries 30.9 \\
        \bottomrule
    \end{tabular}
\end{table}

\section{Conclusion}
\label{sec:outlook}
In this work, we proposed a Teacher-Student approach for the extraction of speaker embeddings. In contrast with the teacher, the student is able to produce sensible speaker embeddings at frame rate.
Further, the student's speaker embeddings contain sufficient information for segmentation and speaker identification to be used as representation of the input speech for neural diarization systems, even during speech overlap.  
Experiments have shown that they help mitigate a  well-known problem of EEND systems: when increasing the number of speakers in a meeting, the performance drop is much less pronounced. Furthermore, we presented a block-wise student-EEND system that can handle arbitrarily long meetings.
Finally, it is worth mentioning that the student embeddings impose no constraints on the backend diarization system. Thus, more recent extensions of EEND-based systems (e.g. EEND-EDA \cite{Horiguchi22EDAEEND}) can be easily combined with the Student-EEND, which serves as an outlook on future work.

\section{Acknowledgement}
Computational resources were provided by the Paderborn Center for Parallel Computing.
Christoph Boeddeker was funded by Deutsche Forschungsgemeinschaft (DFG), project no.\@ 448568305.

\newpage
\bibliographystyle{IEEEbib}
\bibliography{refs}

\end{document}